\documentclass[%
reprint,
superscriptaddress,
amsmath,amssymb,
aps,
prl, 
]{revtex4-2}

\usepackage{graphicx}
\usepackage{subcaption}
\usepackage{dcolumn}
\usepackage{bm}
\usepackage{bbm}
\usepackage[utf8]{inputenc}
\usepackage[english]{babel}

\usepackage[labelformat=simple]{subcaption}
\DeclareMathOperator*{\argmax}{arg\,max}

\begin{document}
	
	\preprint{APS/123-QED}
	
	\title{Robust universal approach to identify travelling chimeras and synchronized clusters in spiking networks}
	
	\author{Olesia Dogonasheva}
	\affiliation{Centre for Cognition and Decision Making, National Research University Higher School of Economics, Moscow, Russia}

	\author{Dmitry Kasatkin}
	\affiliation{Department of Nonlinear dynamics, Institute of Applied Physics RAS, Nizhny Novgorod, Russia}

	\author{Boris Gutkin}
	\affiliation{Centre for Cognition and Decision Making, National Research University Higher School of Economics, Moscow, Russia}
	\affiliation{Group of Neural Theory and LNC2 INSERM U960, École Normale Supérieure PSL* University, Paris, France}

	\author{Denis Zakharov}
	\email{dgzakharov@hse.ru}
	\affiliation{Centre for Cognition and Decision Making, National Research University Higher School of Economics, Moscow, Russia}
	
	
	
	
	
	\begin{abstract}
		We propose a robust universal approach to identify multiple dynamical states, including stationary and travelling chimera states based on an adaptive coherence measure. 
		Our approach allows automatic disambiguation of synchronized clusters, travelling waves, chimera states, and asynchronous regimes. In addition, our method can determine the number of clusters in the case of cluster synchronization. We further couple our approach with a new speed calculation method for travelling chimeras. We validate our approach by an example of a ring network of type II Morris-Lecar neurons with asymmetrical nonlocal inhibitory connections where we identify a rich repertoire of coherent and wave states. We propose that the method is robust for the networks of phase oscillators and extends to a general class of relaxation oscillator networks.
		
	\end{abstract}
	
	\maketitle
	
	
	
\section{Introduction}
	
Chimera states in homogeneous networks, characterized by heterogeneous spatio-temporal patterns containing both synchronous and asynchronous activity regions, have been a concerted focus of modern nonlinear science and synchronization theory \cite{parastesh2020chimeras}. Since their recent dyscovery \cite{abrams2004chimera} they have been identified in multiple systems, such as phase oscillator networks (e.g. \cite{panaggio2015chimera}), the Stuart-Landau systems \cite{zakharova2014chimera, laing2010chimeras}, and found experimentally in multiple systems \cite{hagerstrom2012experimental, tinsley2012chimera, martens2013chimera, viktorov2014coherence, larger2015laser, uy2019optical}. 
Studies of chimera states in the neuronal networks, despite their interest for computational functional importance, are rapidly developing yet are less numerous (\cite{wang2020brief}). 

In fact, coexistence of synchronous and asynchronous neuronal activity (chimera states) have been found experimentally accross brain regions and experimental paradigms \cite{mukhametov1977interhemispheric, rattenborg2000behavioral, compte2000synaptic, lainscsek2019cortical, andrzejak2016all}. The chimera/multichimera states in the neuronal networks have been mostly studied for the nonlocal ring networks with electrical and chemical synapses as well as different neuronal models (e.g. \cite{sakaguchi2006instability, calim2018chimera, omelchenko2013nonlocal, shepelev2017new, wang2020chimeras, bera2016chimera}; reviewed in \cite{wang2020brief}).

One of the more interesting cases of chimera states are travelling chimeras where the incoherent domain moves across the network. Such states have been recently identified across a wide variety of dynamical systems: Kuramoto networks \cite{xie2014multicluster, omel2019travelling}, mechanical systems \cite{dudkowski2019traveling}, electromagnetic oscillators \cite{parastesh2019traveling}, hierarchical topology systems \cite{hizanidis2015chimera}, and neuronal networks \cite{bera2016imperfect, mishra2017traveling}.
	
Despite the increased interest in chimera states, ability to robustly and automatically identify the structure of such complex spatio-temporal dynamics correctly, remains a key challenge. Arguably, without such methodology, results remain exemplary. 
Over the decades several approaches have been used. The simplest -- is the visual method that uses rasterplots, snapshots of instantaneous distributions of the state variable (e.g. neuronal membrane potentials) and frequency distribution diagrams (e.g. \cite{abrams2004chimera, omelchenko2013nonlocal}). This approach gives the network state for an individual set of network parameters and initial conditions. 
For a more automatic and generalizable parametric search for the chimera states, coherence measures have been proposed, like the Kuramoto order parameter \cite{abrams2004chimera}, strength of incoherence (together with a discontinuity measure) \cite{gopal2014observation}, and the $\chi^2$-parameter \cite{golomb2001mechanisms}. All these methods have significant drawbacks for relaxation systems and, in particular, for neuronal networks. The Kuramoto order parameter calculations require well characterized phases of the network elements; these are difficult to define correctly given the fast-slow relaxation dynamics of the spiking and bursting neurons and phase perturbations caused by neuronal interactions. The strength of incoherence is exquisitely sensitive to two method parameters (the number of bins and the coherence threshold) that are difficult to select correctly, especially when the state space includes of a wide variety of synchronous regimes. The $\chi^2$-parameter is more straightforward to compute, yet it suffer from disadvantages inherent in all of these methods. However, since it also has some drawbacks, which we will discuss below, one cannot automatically determine state space regions distinguishing basic dynamical regimes in neuronal networks.

We introduce a new robust universal approach that allows to identify {\it semi-automatically} chimera/multichimera states, both stationary and travelling, based on an extension of the $\chi^2$-parameter. Our approach is free from the above-mentioned drawbacks and can be used to identify correctly and comprehensively dynamical regimes in neuronal networks: global synchronization, cluster synchronization, travelling waves, chimera states, as well as asynchronous behavior. 

We validate the approach, by giving as a paradigmatic example, calculations of dynamical regimes of a ring network of type II-excitable Morris-Lecar neurons \cite{morris1981voltage} with the asymmetrical nonlocal inhibitory connectivity:

\begin{equation}
	\begin{cases}
		C\dot V_i &= I_{app} - g_{Ca}m^{inf}_i(V_i - E_{Ca}) - \\
		&- g_Kw_i(V_i - E_K) - g_L(V_i - E_L) + I^{syn}_i, 
		\\
		\dot w_i &= \phi(w^{inf}_i - w_i)\cosh\frac{V_i - V_3}{2V_4}
		\\
		w^{inf}_i &= 0.5\Bigl(1 + \tanh\frac{V_i - V_3}{V_4}\Bigr)
		\\
		m^{inf}_i &= 0.5\Bigl(1 + \tanh\frac{V_i - V_1}{V_2}\Bigr)
	\end{cases}
	\label{eq1}
\end{equation}
where $i \in {1..N}$, $N$ is number of neurons. $V_i$ is a membrane potential of $i$-th neuron. $w_i$ and $m_i$ are the fraction of open K$^+$ and Ca$^{2+}$ channels, respectively. $E_K$, $E_{Ca}$, $E_L$ are the reversible potentials for potassium, calcium and leak channels, respectively. $g_K$, $g_{Ca}$, $g_L$ are corresponding conductances. $I_{app}$ is an applied current. $C$ is a membrane capacity. The neurons are of II type of excitability.

$I^{syn}_i$ is a synaptic current supplied to $i$-th neuron. The inhibitory GABA$_A$ ($\gamma$-aminobutyric acid-A) synapses are given by first order kinetics \cite{destexhe1998kinetic}:
\begin{equation}
	\begin{cases}
		I^{syn}_i &= \frac{g_{syn}}{N} \sum^{i+R}_{j=0}{x_{ij}} (V_R - V_i)
		\\
		\dot x_{ij} &= \alpha\frac{1}{1 + \exp(-\frac{V_j - V_{syn}}{K_p})}(1 - x_{ij}) - \beta x_{ij}
	\end{cases}
	\label{eq1-1}
\end{equation}
	
Here $V_R$ is a reversal potential, $V_{syn}$ is a threshold, $K_p$ is the synaptic activation, $g_{syn}$ is a synaptic strength, $r = \frac{R}{N}$ is a connectivity parameter, $R$ is a number of connections, $x_i$ is the proportion of open receptors on the postsynaptic membrane of $i$-th neuron.
The topology of the network is a ring. All connections are unidirectional. 
To study network dynamical regimes, we fix all parameters (Table \ref{tab1}) except the control parameters: the external current $I_{app}$, synaptic strength $g_{syn}$ and connectivity parameter $r$.
	
\begin{table}[h!]
	\begin{tabular}{@{}llllll@{}}
		\hline
		$g_K = 8$ mS/cm$^2$      & $E_K = -80$ mV    & $V_1 = -1.2$ mV & $\alpha = 1.1$\\
		$g_{Ca} = 4.4$ mS/cm$^2$ & $E_{Ca} = 120$ mV & $V_2 = 18$ mV   &  $\beta = 0.19$\\
		$g_L = 2$ mS/cm$^2$      & $E_L = -60$ mV    & $V_3 = 2$ mV    & $N = 500$\\
		$\phi = 1 / 25$ & $C = 20$ $\mu$F/cm$^2$  & $V_4 = 30$ mV  & $K_p = 5$  \\
		$V_{syn} = 2$ mV & $V_R = -60$ mV \\ 
		\hline
	\end{tabular}
	\caption{The fixed neuronal network parameters}
	\label{tab1}
\end{table}
	
Numerical integration was performed (\ref{eq1}-\ref{eq1-1}) using the fixed step Euler method ($100$~$\mu$s). Simulation time was $30$~s.

\section{Identification of travelling chimera states}	
	
\subsection{Adaptive coherence measure}
To estimate coherence of a network, one can use $\chi^2$-parameter \cite{golomb2001mechanisms}:
	
\begin{equation}
	\chi^2 = \frac{\sigma^2_V}{\frac{1}{N}\sum^{N}_{i=1}{\sigma^2_{V_i}}},
	\label{chi2}
\end{equation}
where $\sigma^2_V$ is variance of average membrane potential of the network $V(t) = \frac{1}{N}\sum_{i=1}^{N}V_i(t)$:
\begin{equation}
	\sigma^2_V = \langle V^2(t)\rangle_t - \langle V(t)\rangle_t^2,
\end{equation}
and $\sigma^2_{V_i}$ is variance of membrane potential of the i-th neuron:
\begin{equation}
	\sigma^2_{V_i} = \langle V_i^2(t)\rangle_t - \langle V_i(t)\rangle_t^2.
\end{equation}
	
The parameter $\chi^2$ takes the values in [0,1]: $\chi^2 = 0$ corresponds to asynchronous state and $\chi^2 = 1$ for full synchrony. For the values $0 < \chi^2 < 1$ the chimera states may exist. Compared with the Kuramoto order parameter and the strength of incoherence, the advantage of this parameter is that there is no need to calculate the neuronal phases and fine tune the method's intrinsic parameters. However, it cannot distinguish travelling waves, cluster synchronization and chimera states. To solve this problem, we suggest an extended approach, that assumes the identification coherent states with multiple synchronous clusters firing with the time lags. We introduce an Adaptive Coherence Measure (ACM) that based on the solution of the optimization problem:
\begin{equation}
	R^2 = \max_{\mathbf{\Delta t} = (\Delta t_1, \Delta t_2, .., \Delta t_L )} \chi^2(\{V_i(t - \Delta t_i)\}_{i=1}^N),
	\label{max_problem}
\end{equation}
where $\mathbf{\Delta t} = (\Delta t_1, \Delta t_2, .., \Delta t_L)$ is a vector of the unique time lags. 

An analogous approach was used for the similarity function in \cite{rosenblum1997phase}. 

Direct solution of the equation (\ref{max_problem}) involves multidimensional optimization and may require significant computational capacity. To avoid it, let us align the first spikes of all neurons in the network and in this way find the vector $\mathbf{\Delta t}$. This removes the phase shift between the neuronal clusters. For bursters, it is sufficient to align the first spikes of the bursts. Thus, if $R^2=1$ and $1<L\ll N$, there exists cluster synchronization and quantity of synchronous clusters is equal to $L)$. For example, in the Fig.~\ref{fig:cl_syn} there are two synchronous antiphase clusters, $R^2 = 0.9999\approx 1$ and $L=2$ that verify our approach. For comparison, $\chi^2 = 0.2254$ and does not correctly identify the cluster synchronization regime. Note that in the trivial case, when the network is globally synchronized, both $R^2$ and $\chi ^ 2$ takes the equal value of 1 and $L = 1$ which corresponds to a single synchronous cluster. 

\newsavebox{\pic}
\begin{figure}[h!]\centering
	\sbox{\pic}
	{
		\begin{subfigure}[b]{0.45\linewidth}\centering
			\includegraphics[width=1\linewidth]{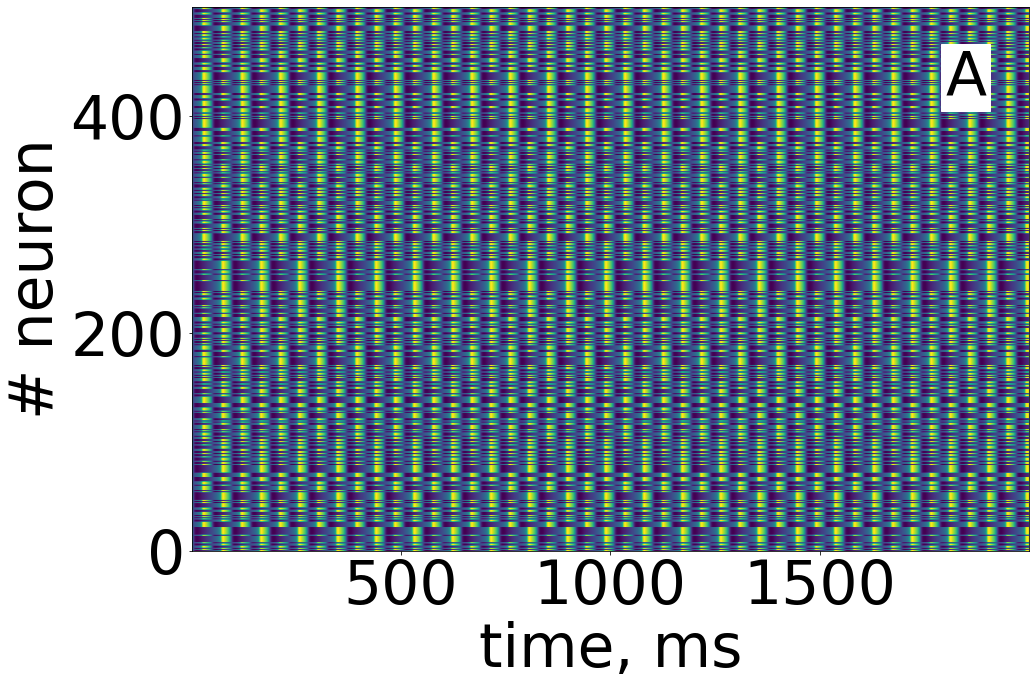}
		\end{subfigure}
	}
	\usebox{\pic}
	\quad 
	\begin{minipage}[b][\ht\pic][s]{0.4\linewidth}
		\begin{center}
			\includegraphics[width=1\linewidth]{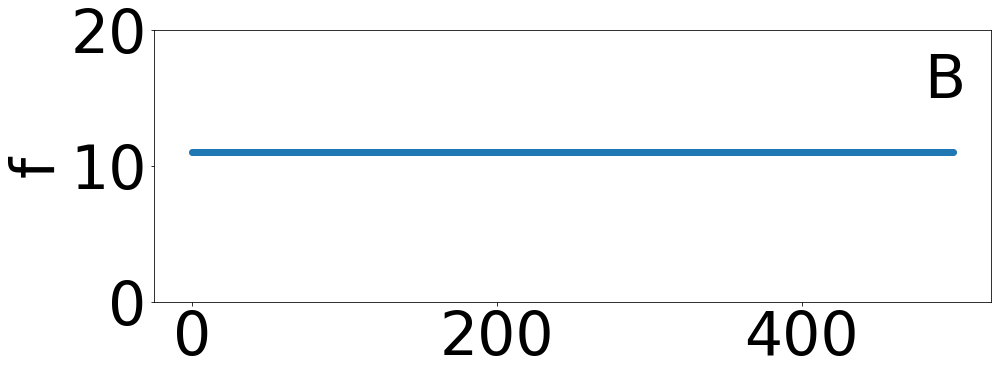}
		\end{center}
		\begin{center}
			\includegraphics[width=1\linewidth]{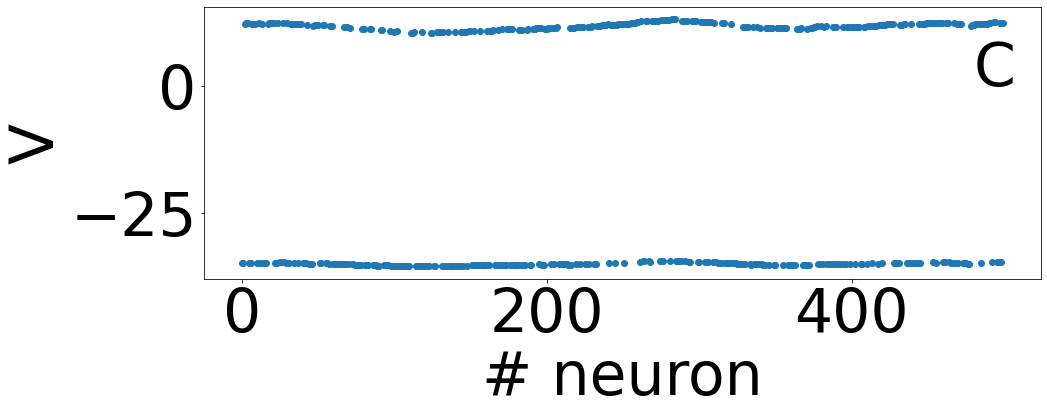}
		\end{center}
	\end{minipage}
	\caption{Rasterplot (A), frequency distribution (B), and snapshot at $t=1000$~ms (C) for the case of two synchronous antiphase clusters ($I_{app} = 95$~$\mu$A/cm$^2$, $g_{syn} = 1$~mS/cm$^2$, $r = 0.9$)}
	\label{fig:cl_syn}
\end{figure}

\begin{figure}[h!]\centering
	\sbox{\pic}
	{
		\begin{subfigure}[b]{0.45\linewidth}\centering
			\includegraphics[width=1\linewidth]{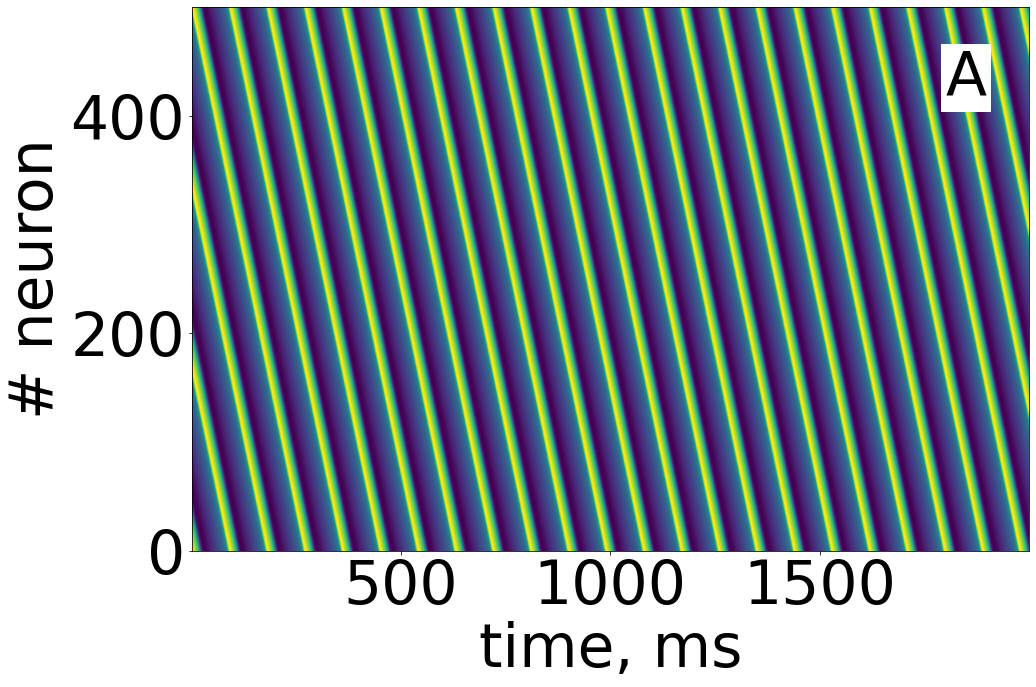}
		\end{subfigure}
	}
	\usebox{\pic}
	\quad 
	\begin{minipage}[b][\ht\pic][s]{0.4\linewidth}
		\begin{center}
			\includegraphics[width=1\linewidth]{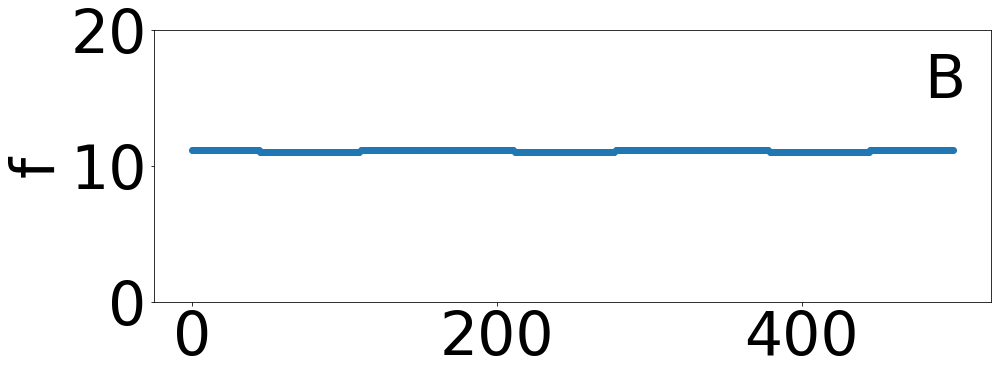}
		\end{center}
		\begin{center}
			\includegraphics[width=1\linewidth]{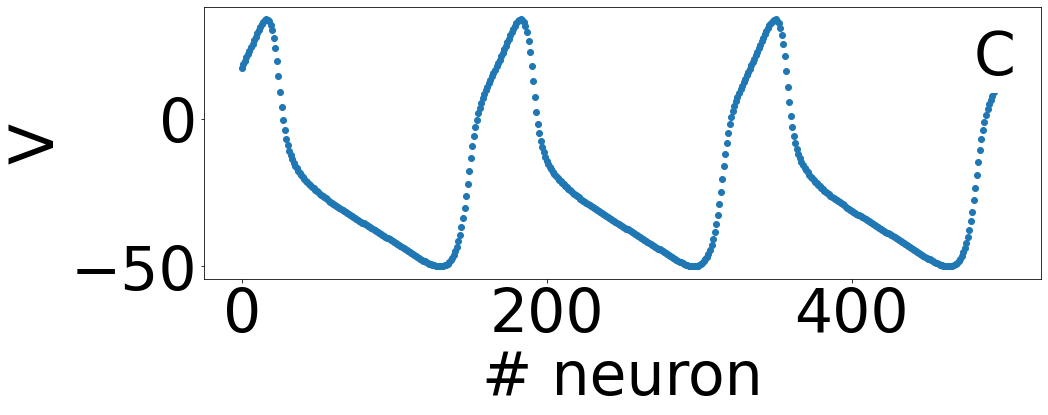}
		\end{center}
	\end{minipage}
	\caption{Rasterplot (A), frequency distribution (B), and snapshot at $t=1000$~ms (C) for a travelling wave ($I_{app} = 95$~$\mu$A/cm$^2$, $g_{syn} = 5$~mS/cm$^2$, $r = 0.2$)}
	\label{fig:tw}
\end{figure}

In the case of travelling wave dynamics (Fig.~\ref{fig:tw}), $\chi^2 = 0.0006$ (like for an asynchronous state) but $R^2=1$ and $L=N$. When the voltage traces are aligned, neurons are fully synchronous, and all of them have a phase shift relative to each other. 
	
\begin{figure}[h!]\centering
	\sbox{\pic}
	{
		\begin{subfigure}[b]{0.45\linewidth}\centering
			\includegraphics[width=1\linewidth]{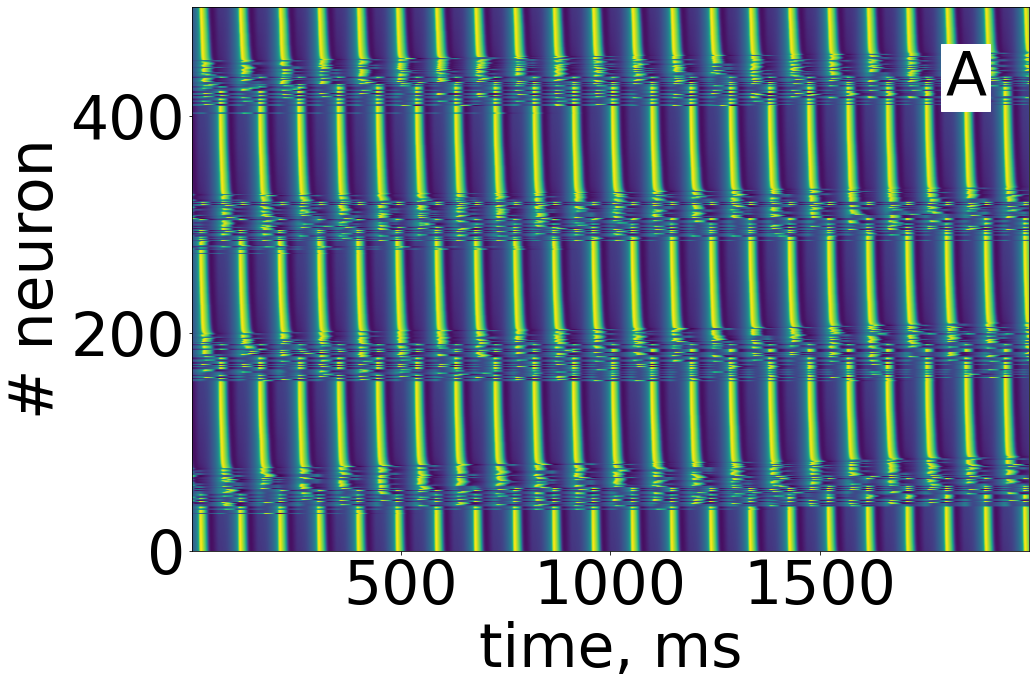}
		\end{subfigure}
	}
	\usebox{\pic}
	\quad 
	\begin{minipage}[b][\ht\pic][s]{0.4\linewidth}
		\begin{center}
			\includegraphics[width=1\linewidth]{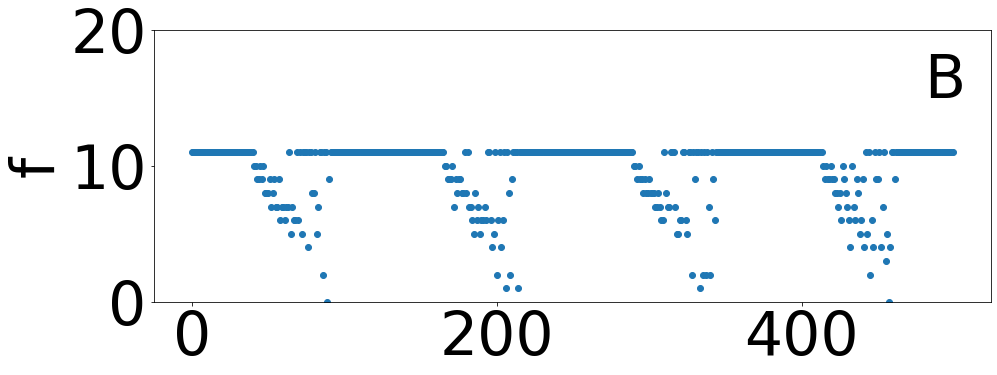}
		\end{center}
		\begin{center}
			\includegraphics[width=1\linewidth]{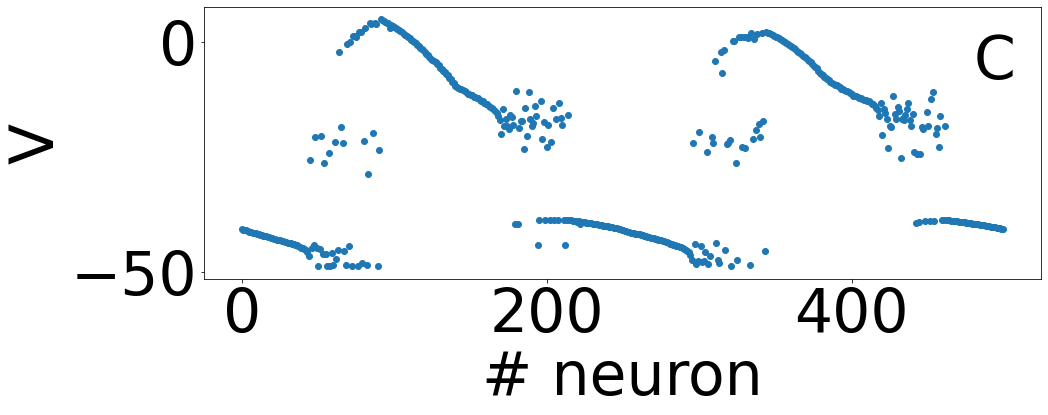}
		\end{center}
	\end{minipage}
	\caption{Rasterplot (A), frequency distribution (B), and snapshot at $t=1000$~ms (C) for the case of a multichimera state ($I_{app} = 95$~$\mu$A/cm$^2$, $g_{syn} = 3$~mS/cm$^2$, $r = 0.84$)}
	\label{fig:mc}
\end{figure}

For a static multichimera state (Fig.~\ref{fig:mc}) both approaches give the values of coherence parameters in $(0,1)$: $\chi^2 = 0.1591$ and $R^2 = 0.8642$. 
Marking the neurons by their time lags, we can determine that there are two large synchronous groups of neurons ($L_{lsg}=2$) and a large population of asynchronous neurons in the network. To make sure that these clusters are synchronous one can calculate the ACM only for the neurons that belong to the clusters. One can distinguish the chimera and multichimera states by searching for alternating coherent and incoherent domains marked by the time lags. 

\begin{figure}[h!]\centering
	\sbox{\pic}
	{
		\begin{subfigure}[b]{0.45\linewidth}\centering
			\includegraphics[width=1\linewidth]{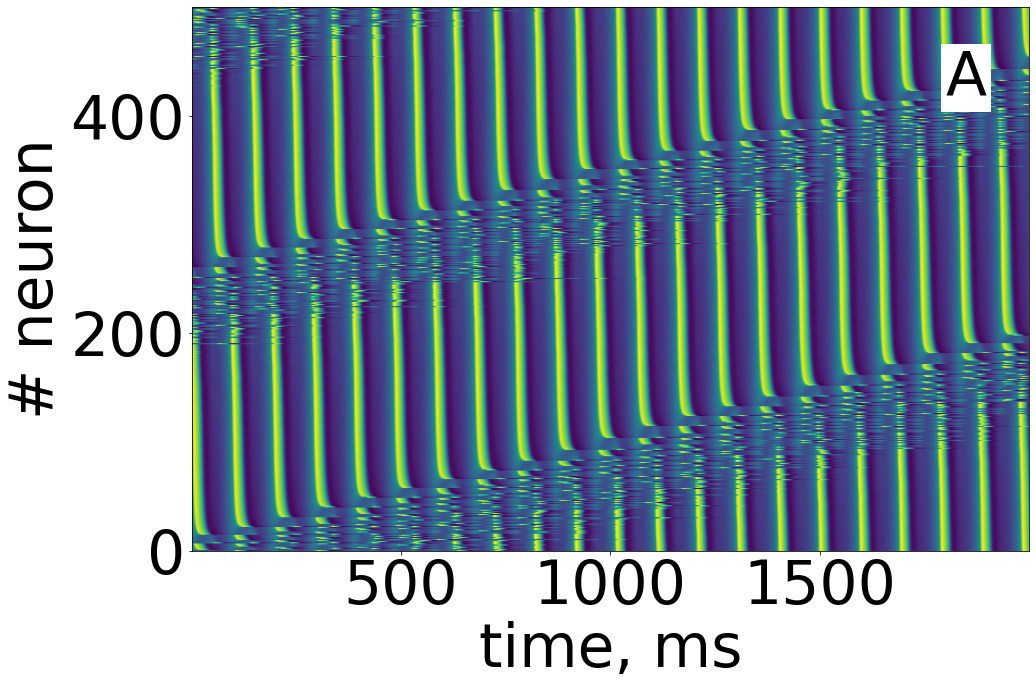}
		\end{subfigure}
	}
	\usebox{\pic}
	\quad 
	\begin{minipage}[b][\ht\pic][s]{0.4\linewidth}
		\begin{center}
			\includegraphics[width=1\linewidth]{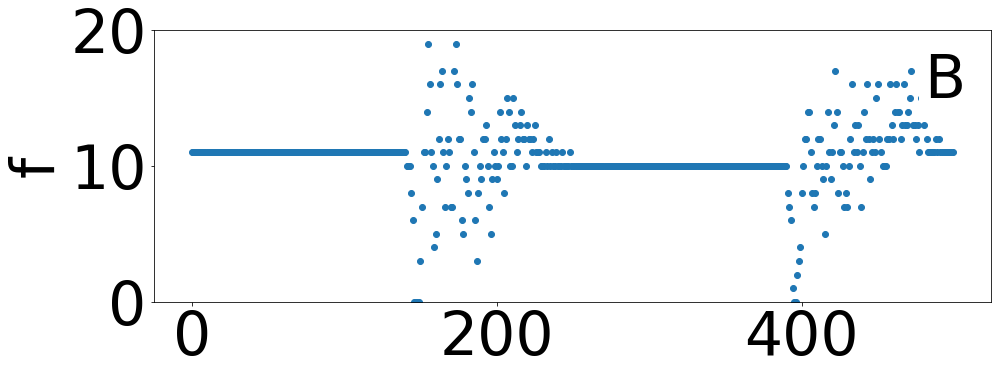}
		\end{center}
		\begin{center}
			\includegraphics[width=1\linewidth]{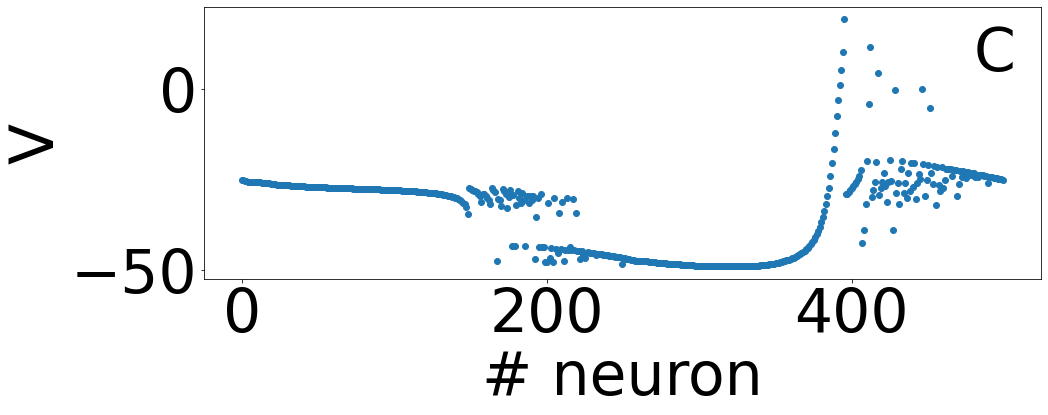}
		\end{center}
	\end{minipage}
	\caption{Rasterplot (A), frequency distribution (B), and snapshot at $t=1000$~ms (C) for the case of a travelling multichimera state ($I_{app} = 95$~$\mu$A/cm$^2$, $g_{syn} = 3$~mS/cm$^2$, $r = 0.78$)}
	\label{fig:t_mc}
\end{figure}

There also exist a travelling chimera state (Fig.~\ref{fig:t_mc}) containing two synchronous clusters for which we can determine the chimera speed $v_{chi} = 0.00878$ neurons per time unit (the method of chimera speed calculation is considered in the next subsection). Changing to the appropriate travelling coordinates, we can ``stop'' the travelling chimera and plot the frequency distribution diagram (Fig.~\ref{fig:t_mc}B). Note that our approach gives an accurate estimation of this chimera state: $R^2 = 0.7779$, $L_{lsg}=2$ in comparison with $\chi^2 = 0.1269$ that improperly corresponds to a low level of coherence.

\begin{figure}[h!]\centering
	\sbox{\pic}
	{
		\begin{subfigure}[b]{0.45\linewidth}\centering
			\includegraphics[width=1\linewidth]{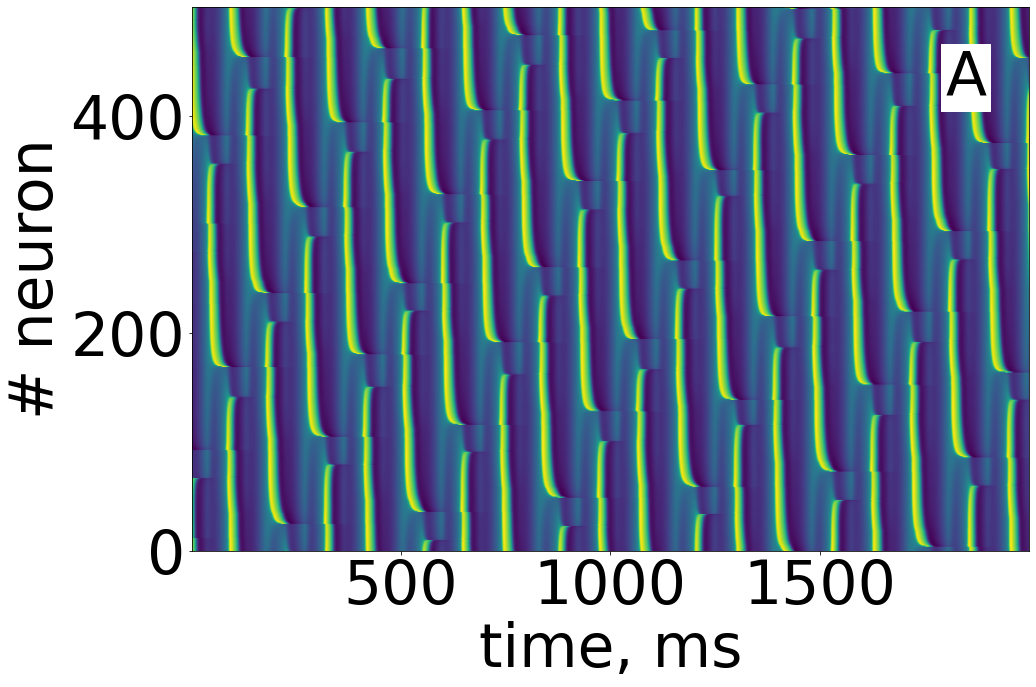}
		\end{subfigure}
	}
	\usebox{\pic}
	\quad 
	\begin{minipage}[b][\ht\pic][s]{0.4\linewidth}
		\begin{center}
			\includegraphics[width=1\linewidth]{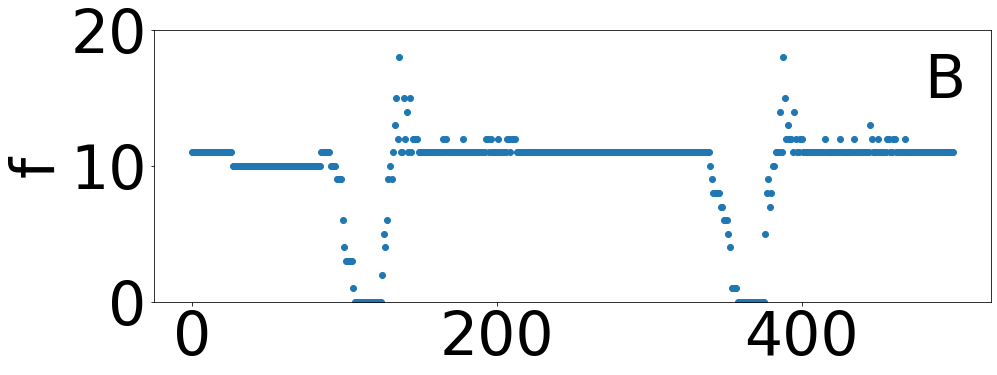}
		\end{center}
		\begin{center}
			\includegraphics[width=1\linewidth]{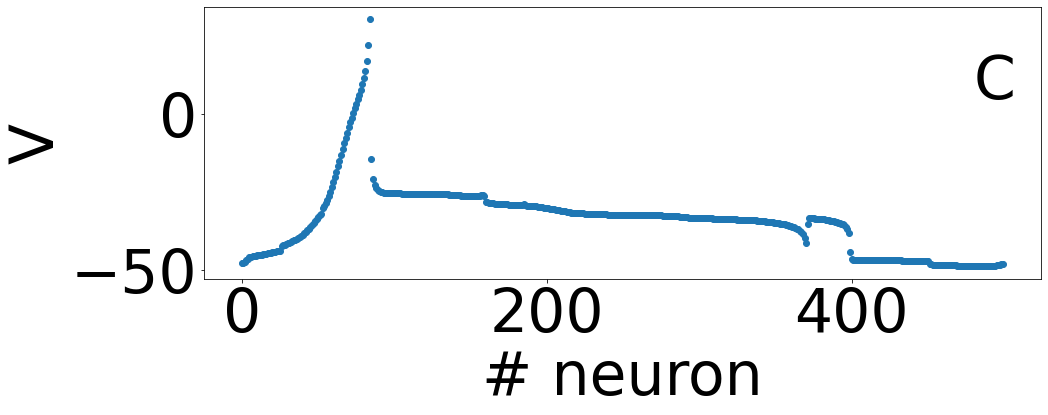}
		\end{center}
	\end{minipage}
	\caption{Rasterplot (A), frequency distribution (B), and snapshot at $t=1000$~ms (C) for the case of the travelling chimera state with domains of subthreshold oscillations ($I_{app} = 95$~$\mu$A/cm$^2$, $g_{syn} = 5.5$~mS/cm$^2$, $r = 0.73$)}
	\label{fig:t_sub}
\end{figure}

Another interesting regime is a travelling chimera state containing domains of synchronous firing and domains of subthreshold oscillations (Fig.~\ref{fig:t_sub}) which has the speed $v_{chi} = 0.07121$ neurons per time unit. As it was for the case of the travelling chimera in Fig.~\ref{fig:t_mc} our coherence parameter ACM correctly characterizes this state $R^2 = 0.5945$ in comparison with $\chi^2 = 0.0908$ that incorrectly shows almost asynchronous behavior.

\begin{table}[h!]
	\begin{tabular}{cccc}
		\hline
		Regime   & ACM   & dimension & number \\
		&& of $\mathbf{\Delta t}$&of clusters\\
		\hline
		Asynchronous &$R^2=0$& $-$  & $-$\\
		state&&&\\
		\hline
		Global  &$R^2=1$& $L=1$  & $L$\\
		synchronization&&&\\
		\hline
		Cluster     & $R^2=1$ & $1<L\ll N$ & $L$\\
		synchronization&&&\\
		\hline
		Travelling wave & $R^2=1$  & $L=N$ & $-$\\
		\hline
		Chimera state & $0<R^2<1$  & - & $L_{lsg}$\\
		\hline
	\end{tabular}
	\caption{Table for parameter values for identification of the different dynamical regimes}
	\label{tab2}
\end{table}

\subsection {Frequency distributions for travelling chimeras and calculation of their speed}		
	
One of the visual criteria to identify a chimera state is a frequency distribution diagram. We define the average spiking frequency (average interspike interval) of the $i$-th neuron as 
\begin{equation}
f[V(i, t)] = \frac{1}{T_{total}} \left|\left\{t: V(i, t) = V_{thr}\text{ } \& \text{ } \dot V(i,t) > 0\right\}\right|,
\label{frequency}
\end{equation}
where $V(i, t)$ is the membrane potential of the $i$-th neuron, $\dot V(i, t)$ is its time derivative, $V_{thr}$ is the value for Poincare section of the neuronal voltage, and $T_{total}$ is the duration of the time series.

\begin{figure}[h!]
	\begin{minipage}[h]{0.8\linewidth}
		\center{\includegraphics[width=1\linewidth]{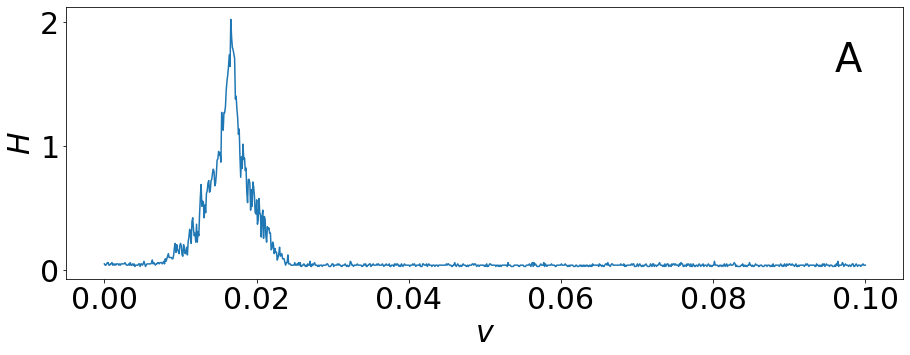}} \\
	\end{minipage}
	\vfill
	\begin{minipage}[h]{0.45\linewidth}
		\center{\includegraphics[width=1\linewidth]{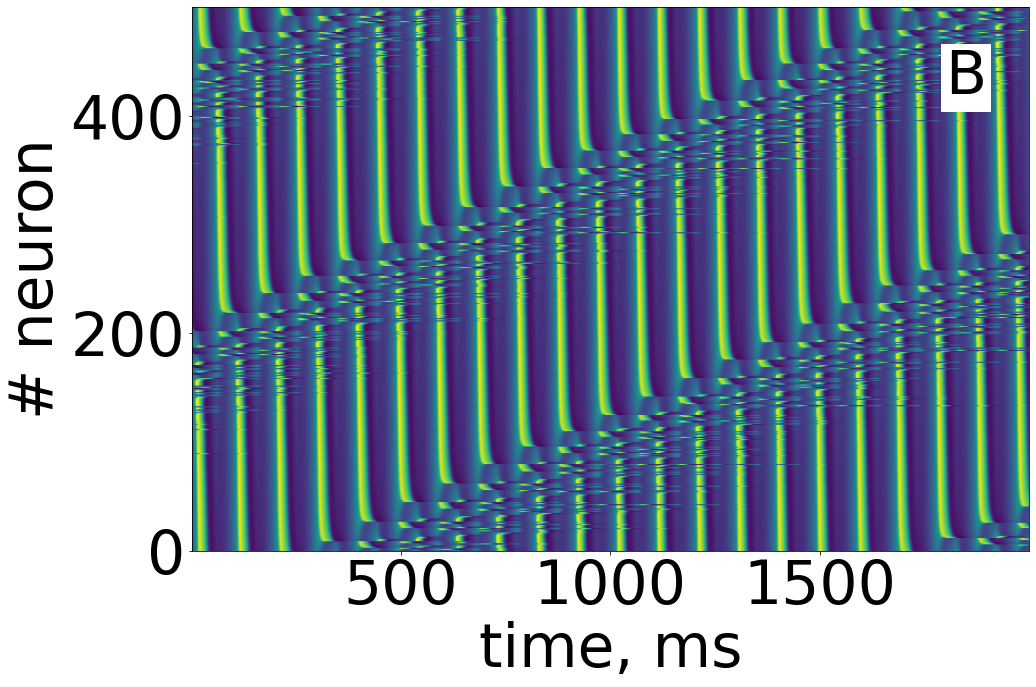}} \\
	\end{minipage}
	\begin{minipage}[h]{0.45\linewidth}
		\center{\includegraphics[width=1\linewidth]{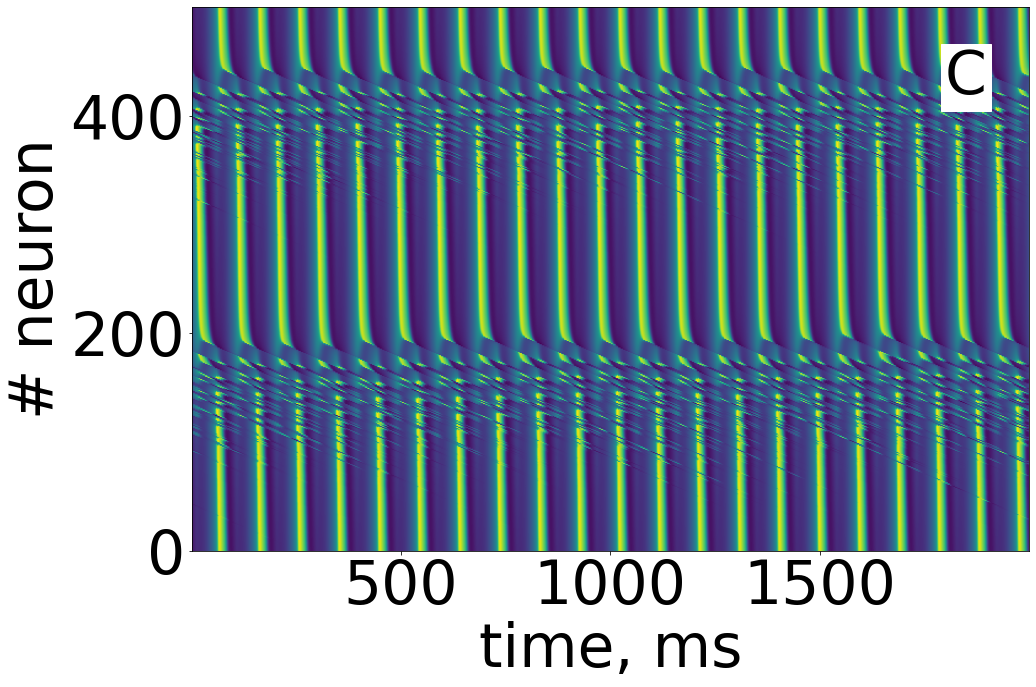}} \\
	\end{minipage}
	\begin{minipage}[h]{0.45\linewidth}
		\center{\includegraphics[width=1\linewidth]{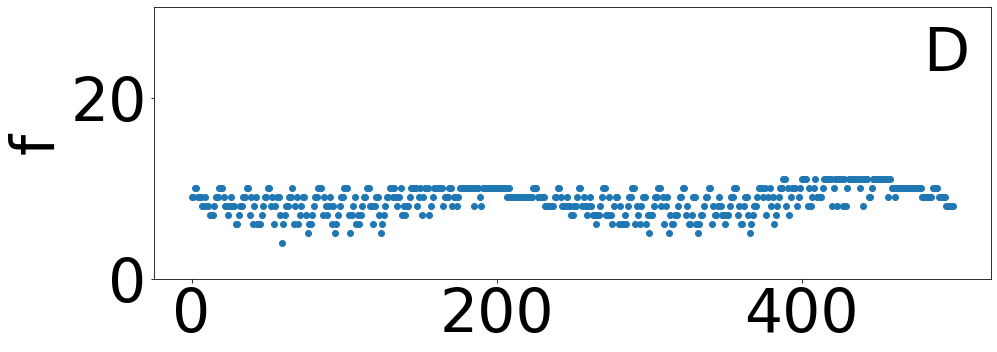}} \\
	\end{minipage}
	\begin{minipage}[h]{0.45\linewidth}
		\center{\includegraphics[width=1\linewidth]{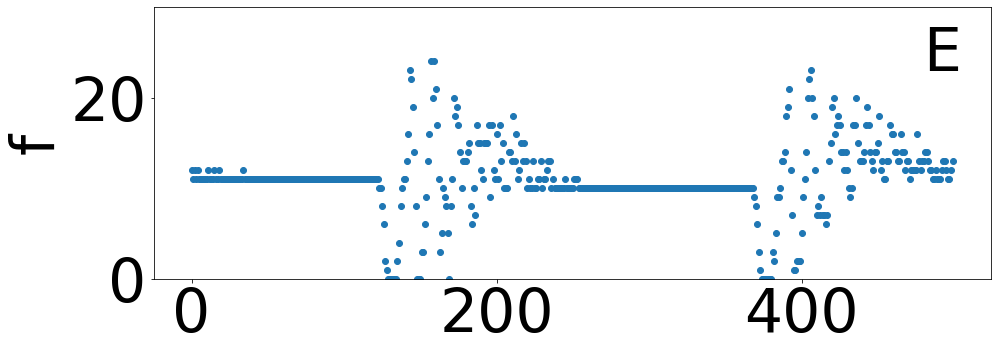}} \\
	\end{minipage}
	\caption{Heuristic functional $H[v]$ for the case of travelling chimera state (A). Initial (B) and traveling chimera coordinates (C) rasterplots and corresponding frequency diagrams (D) and (E). Parameters: $I_{app} = 95$~$\mu$A/cm$^2$, $g_{syn} = 4.5$~mS/cm$^2$, $r = 0.82$}
	\label{stationary_freqs}
\end{figure}

For the travelling chimera state that moves across the ring, it is impossible to plot such diagram directly. Let us define the chimera speed $v_{chi}$ as the number of network elements, the chimera moves per unit of time. Thus, using the travelling wave coordinates ($V(i, t) \rightarrow V(i + v_{chi} t, t)$) we can ``stop'' the chimera (like in Fig.~\ref{stationary_freqs}B,C). In this case, all neurons can be split into the coherent and incoherent domains. In other words, there are some neurons that fire periodically with the same frequencies and the others that show irregular activity (Fig. ~\ref{stationary_freqs}C). Actually, if the number of synchronous neurons is maximized than we can distinctly split the coherent and incoherent chimera domains (Fig.~\ref{stationary_freqs}E) and find properly the value of chimera speed $v_{chi}$. If we also have some clusters of subthreshold oscillations (belong to a chimera like in Fig.~\ref{fig:t_sub}) it would be better to take them into account and also ``align'' them horizontally in the raster plot. According to this idea, let us consider heuristic functional $H[v]$:
\begin{equation}
	H[v] = \frac{N_{coh}[v]}{||N_{coh}[v]||} + \frac{N_{inh}[v]}{||N_{inh}[v]||},
	\label{H}
\end{equation}
which is the weighted sum of the heuristics $N_{coh}[v]$ and $N_{inh}[v]$ that, for the fixed value $v$, show the maximum numbers of neurons oscillating with the same spiking frequency and of neurons demonstrating subthreshold oscillations, respectively. 
Now, one can calculate the chimera speed by solving the optimization problem. In the one-dimensional case it has a form:
\begin{equation}
	v_{chi} = \argmax_v H f[V\left((i + t v)\text{ mod } N, t \right)].
	\label{speed}
\end{equation}

Note that the method works successfully for $N_{coh}[v]$ only, but taking into account inhibited neurons (if they exist) makes it more precise. The example of the typical form of $H[v]$ is shown in Fig.~\ref{stationary_freqs}A. It has a global maximum that can be reliably found by the automated algorithms.

Alternatively, frequency variance can be used as an heuristic functional: 
\begin{equation}
H[v] = D[v] =  \sum_{j = 1}^N\left(f_j[v] - \langle f_i[v]\rangle_i\right)^2
\end{equation}
where $f_i[v] = f[V\left((i + tv)\text{ mod } N, t \right)]$ and $\langle f_i[v]\rangle_i$ is the spatially average. The only drawback of this method is that in some cases a several maxima exist. Therefore, extra effort is needed to select the correct one.

\section{Conclusion}
	
In this paper, we introduced a robust and universal approach for studying synchronization processes in neuronal networks based on the ACM-parameter. In comparison with other methods, it has the following advantages: absence of internal method parameters, simplicity of calculations and the ability to {\it automatically} distinguish between stationary and travelling chimera states, global and cluster synchronization, and travelling waves (Table~\ref{tab2}). In addition, our approach can find the number of synchronous clusters and chimera speed if it travels. 

Of course, there are situations where previously proposed methods are appropriate. For example, if the elements of a network have well defined phases, the Kuramoto order parameter \cite{abrams2004chimera} is highly recommended and may also apply to neural networks under specific circumstances (see, for example, \cite{omelchenko2013nonlocal, omelchenko2015robustness, chouzouris2018chimera}). We also point out that although our approach was created for spiking neuronal networks, it also gives excellent results for multiple network classes, and, in particular, for the networks of phase oscillators (simulations not shown). Therefore, we would suggest that our methodology is more universal and is able to differentiate the cluster and chimera states, as well as determine the number and size of synchronous clusters, and can be very useful in networks where it is necessary to separate these states in the parameter space like \cite{kasatkin2017self}.
	
We should also like to comment that the strength of incoherence \cite{gopal2014observation} together with the discontinuity measure, while being able to differentiate chimera and multichimera states, cannot do so for coherent states consisting of multiple synchronous clusters and chimeras. In this case, to avoid confusion, the authors suggested studying the distribution of membrane potentials differences in the network to eliminate the removable discontinuity. 
Most importantly, this method is exquisitely sensitive to the choice of method parameters (bin number and the coherence threshold). For example, for travelling waves the strength of incoherence may be equal to either 1 or 0 (with a jump between these values) depending on the coherence threshold. Almost same situation is observed for the case of travelling chimeras and it cannot distinguish between cluster synchronization and asynchronous state.

In addition, we suggest the new method of chimera speed calculation that allows to stop the chimera and properly separate coherent and incoherent chimera domains. Also, there exists a method \cite{hizanidis2015chimera} based on Fourier spectra that has been suggested for calculation of the chimera speed for the lattice limit cycle (LLC) model and shown to be useful for neuronal networks (see, for example, \cite{bera2016imperfect}). Our method has some significant advantages. First, it is simpler for implementation and calculations can be easily automated. Second, one does not need to select the ``right'' maximum in the Fourier spectra, but to find the global maximum of $H[v]$. Also, our method is less demanding on the simulation time and, in fact, the neuronal time series should contain only one non-coherent cluster of a chimera. Increasing of a simulation time only leads to a slight growth in accuracy.

To sum up, we propose a universal and robust methodology to characterize multiple complex spatio-temporal patterns in a wide range of coupled networks, notably those with relaxation dynamics of the constituent elements.

\section{Acknowledgements}
	
The research leading to these results has received funding from the Basic Research Program at the National Research University Higher School of Economics. This research was supported in part through computational resources of HPC facilities at NRU HSE.
BG acknowledge support from the ANR Project ERMUNDY (Grant No ANR-18-CE37-0014)
	
\bibliography{apssamp}
	
\end{document}